   \documentclass[twocolumn            ]{aastex631}
\hypersetup{linkcolor=blue,citecolor=blue,filecolor=blue,urlcolor=blue   }
\shorttitle{Sgr A* position and proper motion}
\shortauthors{Gordon et al.}
\begin{document}

\title{Position and Proper Motion of Sagittarius A* in the ICRF3 Frame from 
VLBI Absolute Astrometry}

\correspondingauthor{David Gordon}
\email{david.gordon126.civ@us.navy.mil}

\author[0000-0001-8009-995X]{David Gordon}
\affil{U.S. Naval Observatory     \\
3450 Massachusetts Ave NW, Washington, DC 20392, USA}

\author[0000-0001-9885-4220]{Aletha de Witt}
\affiliation{South African Radio Astronomy Observatory \\
PO Box 443, Krugersdorp 1740, South Africa }

\author[0000-0003-0828-1401]{Christopher S. Jacobs}
\affiliation{Jet Propulsion Laboratory, California Institute of Technology/NASA \\
4800 Oak Grove Drive, Pasadena, California 91109, USA }



\begin{abstract}
\
Sagittarius A* (Sgr A*) is a strong, compact radio source believed to
be powered by a super-massive black hole at the galactic center. Extinction 
by dust and gas in the galactic plane prevents observing it optically, but 
its position and proper motion have previously been estimated using radio 
interferometry. We present new VLBI absolute astrometry measurements of its 
precise position and proper motion in the frame of the third realization of 
the International Celestial Reference Frame, ICRF3. The observations used 
were made at 52 epochs on the VLBA at K-band (24 GHz) between June 
2006 and August 2022. 
We find the proper motion of Sgr A* to be -3.128 $\pm$ 0.042 
mas ${\rm yr}^{-1}$ in right ascension and -5.584 $\pm$ 0.075 mas ${\rm yr}^{-1}$ 
in declination, or 6.400 $\pm$ 0.073 mas ${\rm yr}^{-1}$
at a position angle of 209.26$\degr$ $\pm$ 0.51$\degr$.
We also find its J2000 ICRF3 coordinates at the 2015.0 proper motion epoch to be
17$^h$45$^m$40$\fs$034047 $\pm$ 0$\fs$000018,
-29$\degr$00$\arcmin$28$\farcs$21601 $\pm$ 0$\farcs$00044.
In galactic coordinates, Sgr A* shows proper motion of -6.396 $\pm$ 0.071
mas ${\rm yr}^{-1}$ in galactic longitude and -0.239 $\pm$ 0.045 mas ${\rm yr}^{-1}$ in galactic
latitude, indicating solar motion of 248.0 $\pm$ 2.8 km ${\rm s}^{-1}$ in the galactic 
plane and 9.3 $\pm$ 1.9 km ${\rm s}^{-1}$ towards the north galactic pole.
  
\end{abstract}




\section{Introduction} \label{sec:intro}

Accurately locating the galactic center and measuring its proper motion
is important for defining the galactic coordinate system and  
for studies of galactic structure, kinematics and dynamics and for
identification with nearby sources in the radio and IR.
Sagittarius A* (hereafter Sgr A*) has been shown to be a super-massive 
black hole at the galaxy's center (\citet{ghez2008, genzel2010}). 
Previous estimates of its position using the VLA and the VLBA
have been made by \citet{yusef1999}, 
\citet{reid2004}, and \citet{petrov2011}. Previous estimates 
of its proper motion have been made using VLBI phase referencing
(relative, or differential astrometry) by 
\citet{backer1999} on the VLA, and by \citet{reid2004, reid2020} on the VLBA.
In this paper, we present the results of K-band (24 GHz) VLBA observations
in the absolute astrometry mode over a span of 16 years,
leading to a precise determination of its position with respect to time
in the frame of the third realization of the International Celestial
Reference Frame, ICRF3 \citep{charlot2020}.

An initial K-band astrometric catalog of 268 sources observed on 
the VLBA from 2002--2007 was published by \citet{lanyi2010}.
In 2014 this program was resumed by the authors and others with VLBI 
observations in both the northern and southern hemispheres in
anticipation of including a K-band catalog in the release of the third 
realization of the International Celestial Reference Frame, ICRF3
\citep{charlot2020}. ICRF3 was adopted by the IAU in January 2019 and 
contained celestial reference frame catalogs at three frequencies, the
largest at the traditional geodetic/astrometric X/S bands (8.6/2.3 GHz)
and smaller catalogs at K (24 GHz) and X/Ka (8.4/32 GHz) bands. Though not 
included in the ICRF3-K catalog because it is not an extragalactic source, 
Sgr A* was observed and detected in several of the second era K-band 
VLBA sessions beginning in November 2017. K-band observations have
continued into the present time and there is now nearly four
times as much K-band astrometry data as was used for ICRF3. 

Sgr A* is not visible optically due to extinction by the gas and dust in
the galactic plane, but it is a strong radio source.
However, scattering by the ionized interstellar
medium broadens it out to apparent angular sizes 
approximately proportional to the square of the wavelength 
\citep{davies1976,johnson2018}. This broadening 
causes it to be resolved out in lower frequency VLBI observations. 
K-band is the only one of the three ICRF3 frequency bands that can practically
observe Sgr A*. It is not detectable at all in the S-band of the traditional 
X/S geodetic/astrometric observations and only weakly detectable in the X-band, 
where it is smeared out over $\sim$9 mas (FWHM). It also gets resolved out at 
both the X and Ka frequencies in the very long, single-baseline observations 
used to construct the ICRF3-X/Ka catalog. Even at K-band, it is
detectable only on the shorter continental VLBA baselines and never on the 
longer baselines to MK-VLBA and SC-VLBA.

\section{Observations} \label{sec:observations}

Our K-band VLBI Sgr A* data begins with two sessions in June 2006 from the 
original processing at the Goddard Spaceflight Center of the VGaPS 
sessions (VLBA sessions BP125A and BP125B, described by \citet{petrov2011}, 
which were recorded at 256 Mbps). Under the new campaign, we have re-observed 
Sgr A* on the VLBA at 50 additional epochs between November 2017 
and August 2022 as part of observations for ICRF3 
and subsequent maintenance and expansion. These sessions were made in the absolute
astrometry mode (also sometimes called the global astrometry mode) in 
which dozens or hundreds of sources spread out over the entire 
sky are observed in each session in order to determine their positions
in an inertial coordinate system. In each of these 24-hour sessions, \mbox{Sgr A*} 
was observed along with $\sim$225 compact extragalactic ICRF3 sources, thus 
providing a strong link to the ICRF3 reference frame. For the first 19 epochs 
of the revised campaign (November 2017 through September 2019) the right 
circular polarization (RCP) signal was recorded at a bit rate of 2 Gbps. For 
the last 31 epochs, beginning in November 2019, both the RCP and LCP 
signals were recorded for a total of 4 Gbps, and processed 
into separate RCP and LCP databases. Further details of the observations 
are given in \citet{charlot2020}.
Each session was approximately 24 hours in duration and \mbox{Sgr A*} was 
observed in either 2 or 3 90-second scans,
taking up only $\sim$0.5\% or less of the observing time. It was included 
primarily to locate it precisely in the ICRF3 frame. A secondary result is
that these observations provide a time series of the positions of Sgr A*,
which we have analyzed to determine an absolute
astrometric measurement of its proper motion, which can be directly compared to 
several VLBI relative astrometry (phase referencing) measurements.

\section{Analysis} \label{sec:analysis}

The data for Sgr A* was taken from an astrometric solution of 165
K-band databases 
spanning May 2002 - August 2022, generated as an unofficial
update of the \mbox{ICRF3-K} catalog.\footnote{See
https://crf.usno.navy.mil/quarterly-vlbi-solution for the latest global
source solution.}
As in ICRF3, a galactic aberration model 
\citep{macmillan2019} using a galactic aberration constant of 5.8 $\mu$as ${\rm yr}^{-1}$
was applied in this solution, giving J2000 source positions at the galactic 
aberration epoch of 2015.0. However, the galactic aberration model
has no effect in the direction of the galactic center, so this need not be
discussed further.

Correlation of the VLBA K-band data used in this study, except for the 2006 
data, was done on the VLBA's DiFX software correlator \citep{deller2011} at the 
National Radio Astronomy Observatory in Socorro, 
New Mexico. The DiFX output was converted to both Haystack Mark4 format for 
astrometry processing and FITS-IDI format for imaging \citep{deWitt2022}.
Fringe fitting to obtain the geodetic observables (group delays and phase
delay rates) was performed at either the U.S. Naval Observatory, the Goddard 
Space Flight Center or the Bonn correlator, using the Haystack Observatory 
\textit{hops} software package \citep{rogers1995} and the output was
converted into geodetic-style VLBI databases. Astrometric analysis of each 
session was made at either the U.S. Naval Observatory (USNO) or at the Goddard 
Space Flight Center using the \textit{Calc/Solve} \citep{ma1986,charlot2020} 
and \textit{nuSolve} \citep{bolotin2014} analysis packages.
Delays due to the ionosphere were computed as described in \citet{lanyi2010}
and \citet{charlot2020}, using global navigation satellite systems (GNSS) 
ionosphere maps produced by the Jet Propulsion Laboratory. These 
maps are given at 2-hour intervals and have a resolution of 2.5$\degr$ by 
5$\degr$ in latitude and longitude, respectively. Although this temporal and 
spatial resolution does not provide the accuracy of simultaneous dual-frequency 
observations, it does allow the averaging out of most of the systematic effects 
of the ionosphere.

\begin{figure}[ht!] 
       \centering
       \includegraphics[width=.45\textwidth]
%       {SGR-A.UD009AH.colormap.pdf}
        {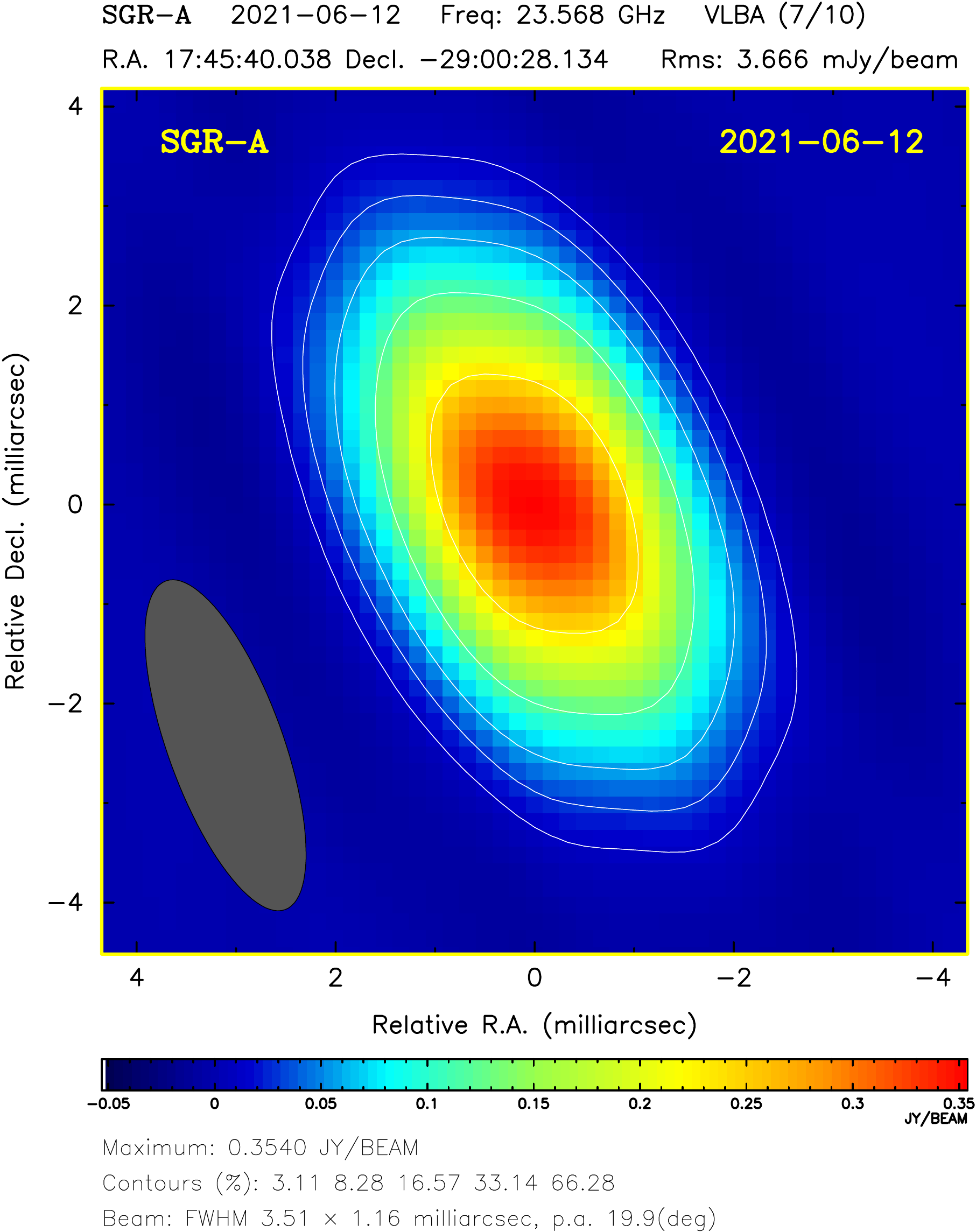}
        \caption {A VLBA image of Sgr A* at K-band, from observations on
	2021 June 12. Note that the ISM broadens the source out, limiting 
	detections to baselines less that $\sim$200 $M\lambda$ ($\sim$2500
	km) at K-band.}
        \label{fig:Kimage}
\end{figure}

Imaging of ICRF3 sources from the K-band VLBA sessions is being done by
\citet{deWitt2022}. A K-band image of Sgr A* based on that work is shown in 
Figure~\ref{fig:Kimage}. 
The estimated FWHM size of Sgr A* at K-band obtained from model fitting is
1.74 milli-arc-sec (mas).
Broadening by the ISM limits detections at K-band to 
baselines of $\sim$200 $M\lambda$  ($\sim$2500 km) or 
%baselines of 200 M\lambda  (2500 km) or 
less. Thus we do not get detections on any of the baselines to MK-VLBA or 
SC-VLBA, and only on a few of the baselines to BR-VLBA and HN-VLBA. This 
results in fewer observations and larger single session uncertainties than for 
most of the extragalactic sources in the sessions. The single sessions each 
get an average of $\sim$25 baseline observations of Sgr A* and have position 
uncertainties mostly in the range of 0.5--1.5 mas. 
\\
   
\section{Astrometric Results} \label{sec:results}

\begin{figure}[h!] 
       \centering
       \includegraphics[width=.45\textwidth]
       {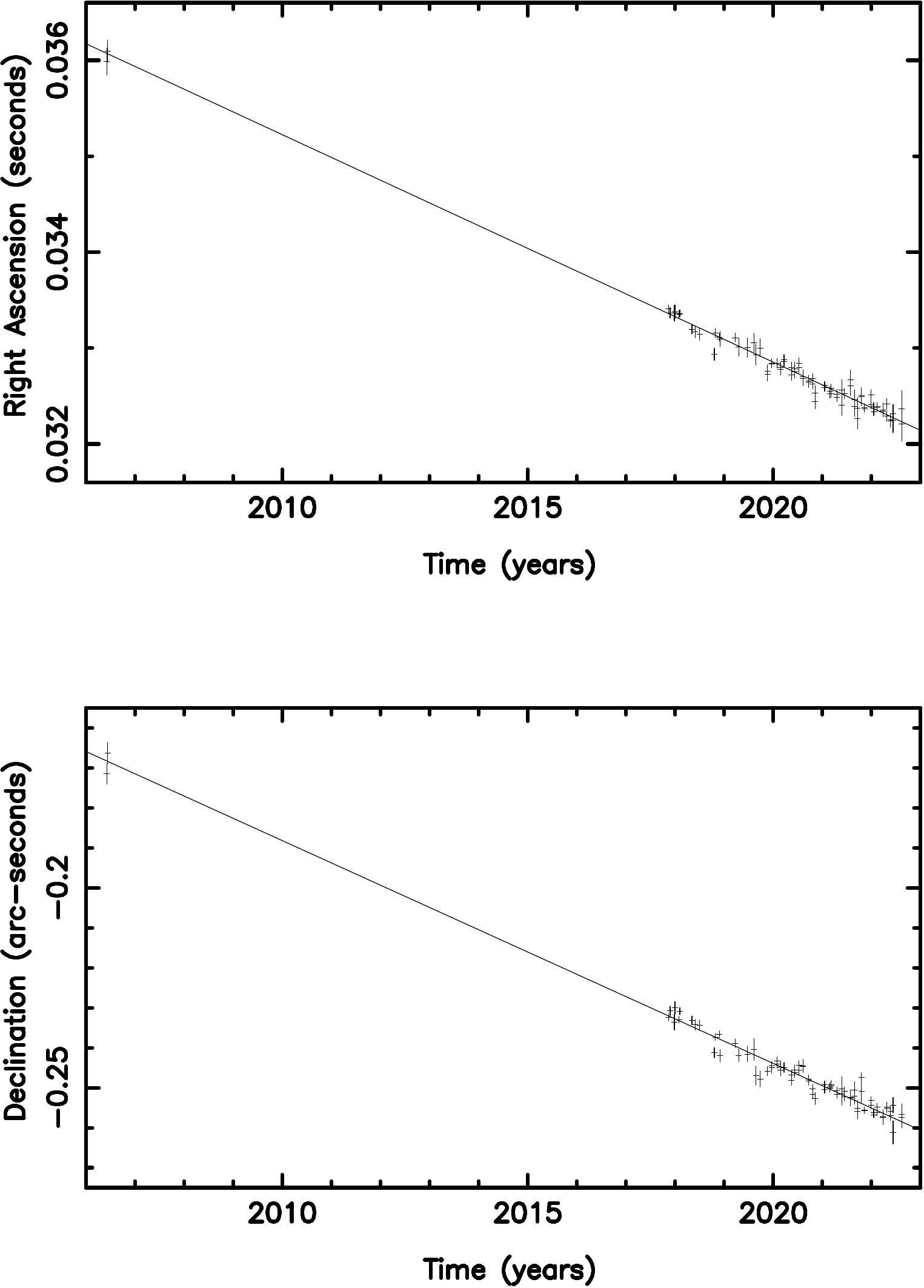}
       \caption {RA's and Dec's of Sgr A* versus time as measured in
       K-band VLBA sessions from June 2006 - August 2022, offset from
       17$^h$45$^m$40$^s, -29$\degr$00$\arcmin$28\arcsec$. WRMS deviations
       from the linear fits are 0.76 mas in RA and 1.71 mas in Dec.}
       \label{fig:sgrA}
\end{figure}

Two methods were used to solve for the position and proper motion of Sgr A*
using the \textit{Calc/Solve} analysis package. In one, 1035 extragalactic 
ICRF sources were treated as global parameters, giving single average J2000 
positions for each extragalactic source, whereas 
the position of Sgr A* was solved for individually in each session, giving
a time series of its J2000 positions. From this time series, we solved for
its proper motion in right ascension (RA) and declination (Dec) and its 
position at the 2015.0 proper motion epoch. In the second solution, both its 
position and proper
motion were solved for globally within \textit{Solve}. 
Alignment with ICRF3 in both solutions was accomplished through a 
no-net-rotation constraint on 258 of the ICRF3 defining sources that have 
been well-observed at K-band. Results of the two methods differed by much 
less than the \mbox{1-sigma} uncertainties. The time series positions are
useful for plotting and examining the data, however, the
global solution incorporates more of the 
correlations between the various parameters and can thus be more accurate. 
A plot of the RA's and Dec's
versus time from the time series solution is shown in Figure~\ref{fig:sgrA}. 
The global solution finds the ICRF3 J2000 position of Sgr A* at the
2015.0 proper motion epoch to be: \\
\indent  $\alpha$ =  ~17$^h$45$^m$40$\fs$034047 $\pm$ 0$\fs$000018 \\
\indent  $\delta$ = -29$\degr$00$\arcmin$28$\farcs$21601~~~ $\pm$ 0$\farcs$00044 \\
The 2015.0 epoch was chosen for consistency with the ICRF3 galactic aberration
epoch \citep{charlot2020}.
The global solution also finds the proper motions in RA and Dec to be:   \\
\indent   $\mu_{\alpha}\cos{\delta}$ = {-3.128 $\pm$ 0.042} mas ${\rm yr}^{-1}$  \\
\indent   $\mu_{\delta}$ = {-5.584 $\pm$ 0.075} mas ${\rm yr}^{-1}$    \\
with a correlation coefficient of 0.2102 between the RA and Dec proper 
motions.
Thus, combined with its proper motion, the J2000 position of Sgr A* 
as a function of time in the ICRF3 frame is: \\
\indent $\alpha$ = ~17$^h$45$^m$40$\fs$034047 $-$ (0$\fs$000238408)$\times$(t - 2015.0)  \\
\indent $\delta$ = -29$\degr$00$\arcmin$28$\farcs$21601~~ $-$ ~ (0$\farcs$005584)$\times$(t - 2015.0)  \\
\indent ~~~~where t is the desired epoch in years. \\
Solving for an annual parallax was not attempted since the expected 
parallax of $\sim$120 $\mu$as would be considerably smaller than the 
individual position uncertainties.

In Table \ref{tab:table1}, we compare our ICRF3 position at 2015.0 
with 3 previous estimates.
For \citet{yusef1999} and \citet{petrov2011}, we extrapolate their 
positions to 2015.0 using our proper motions. 
For \citet{reid2004} we use their proper motions to extrapolate to 2015.0, 
assuming their epoch is at the midpoint of their observations (i.e. 1999.25.)
The two phase referencing studies differ from our ICRF3 position 
at about the 25 and 40 mas level while the VGaPS absolute astrometry position 
differs at about the 2 mas level, although with much lower precision than our
result.

\begin{table}[h!]
\begin{center}
% \label{tab:table1}
 \begin{tabular}{|l|r|r|r|}
 
\hline
            & RA @ 2015.0~~~   &  Dec @ 2015.0~~ & Diff (mas) \\
\hline
This        & 17$^h$45$^m$40$\fs$034047  & -29$\degr$00$\arcmin$28$\farcs$21601 &      \\
work        & $\pm$ 0$\fs$000018~~~~~   & $\pm$ 0$\farcs$00044~~~~~~    &            \\
\hline
Y-Z         & 17$^h$45$^m$40$\fs$0329~~~~   & -29$\degr$00$\arcmin$28$\farcs$195 ~~  & 15.0/21.0~  \\
            & $\pm$ 0$\fs$0007~~~~~~     & $\pm$ 0$\farcs$014~~~~~~     &            \\
\hline
R-B         & 17$^h$45$^m$40$\fs$0371~~  & -29$\degr$00$\arcmin$28$\farcs$207~~  & 40.1/8.8~  \\
\hline
P-K         & 17$^h$45$^m$40$\fs$033943~  & -29$\degr$00$\arcmin$28$\farcs$2177  & 1.4/1.7~~  \\
            &   $\pm$ 0$\fs$00011~~~~~   &    $\pm$ $\farcs$0027~~~~~~     &            \\
\hline
\end{tabular}
\end{center}
\caption{Positions of Sgr A* in the J2000 frame at proper motion epoch 2015.0 
from this study compared to
\citet{yusef1999} (Y-Z), \citet{reid2004} (R-B), and \citet{petrov2011} (P-K).}       
\label{tab:table1}
\end{table}

Combining our RA and Dec proper motions gives a total proper motion of
6.400 $\pm$ 0.073 mas ${\rm yr}^{-1}$ at a position angle of 209.26$\degr$ 
$\pm$ 0.51$\degr$. In Table \ref{tab:table2}, we
compare the RA, Dec, and total proper motions from this study 
to those from \citet{reid2020}, \citet{reid2004} and
\citet{backer1999}. Our absolute astrometry proper motion value
agrees well with the relative astrometry value of \citet{reid2020}, 
but with larger uncertainties for several reasons.
\citet{reid2020} observed 
at Q-band (43 GHz), where the apparent angular size should be only $\sim$1/3
as large as at K-band and they observed at 23 epochs over an 18 year span, 
from 1995 - 2013, with each 8-hour session dedicated to Sgr A* and to two
nearby calibrator sources. Our data on the other hand is from 52 epochs
spanning nearly 16 years, but with all but two epochs spanning only 4.75
years, and with only 3 or 4.5 minutes observing time at each epoch.
The two methods should be considered complementary though. The absolute 
astrometry of this study can precisely locate Sgr A* in the IAU's 
official reference frame, ICRF3, whereas the  relative astrometry cannot 
since there are no nearby ICRF3 sources to use as phase referencing calibrators.
The relative astrometry can yield smaller uncertainties in proper motion
though, due to the very small angles being measured between the target and the 
calibrators, which differentially cancels out most error sources. 
Relative astrometry however can be subject to systematic 
errors due to small shifts in the effective calibrator source positions
over time as a result of source structure or other effects. Such effects would 
mostly average out though over the 258 ICRF3 defining sources used 
to define the reference frame in our absolute astrometry study.
   
%%%%%%%%

\begin{table}[htb!]
\begin{center}
% \begin{tabular}{|l|r|r|r|}
  \begin{tabular}{|l|c|c|c|}
 
\hline
            & $\mu_{\alpha}\cos{\delta}$  & $\mu_{\delta}$         & Combined           \\
\hline
This        & -3.128$\pm$0.042       & -5.584$\pm$0.042      & 6.400$\pm$0.073 mas ${\rm yr}^{-1}$  \\
work        & mas ${\rm yr}^{-1}$      &   mas ${\rm yr}^{-1}$   & @ 209.26$\degr$             \\                
\hline
 R-B2       & -3.156$\pm$0.006       & -5.585$\pm$0.010      & 6.415$\pm$0.008 mas ${\rm yr}^{-1}$ \\
            &  mas ${\rm yr}^{-1}$     &  mas ${\rm yr}^{-1}$    &  @ 209.47$\degr$         \\
\hline
 R-B1       & -3.151$\pm$0.018       & -5.547$\pm$0.026      & 6.379$\pm$0.024 mas ${\rm yr}^{-1}$ \\
            &  mas ${\rm yr}^{-1}$     &  mas ${\rm yr}^{-1}$    &   @ 209.60$\degr$         \\
\hline
 B-S        & -2.70$\pm$0.15         & -5.60$\pm$0.20        & 6.22$\pm$0.25 mas ${\rm yr}^{-1}$    \\
            &  mas ${\rm yr}^{-1}$     & mas ${\rm yr}^{-1}$     & @ 205.7$\degr$         \\
\hline

\end{tabular}
\end{center}
\caption{Proper motion measurements in RA and Dec of Sgr A* 
from this study and from \citet{reid2020} (R-B2),
 \citet{reid2004} (R-B1) and \citet{backer1999} (B-S).}
\label{tab:table2}
\end{table}

\begin{figure*}[ht!] 
       \centering
       \includegraphics[width=1.00\textwidth]
       {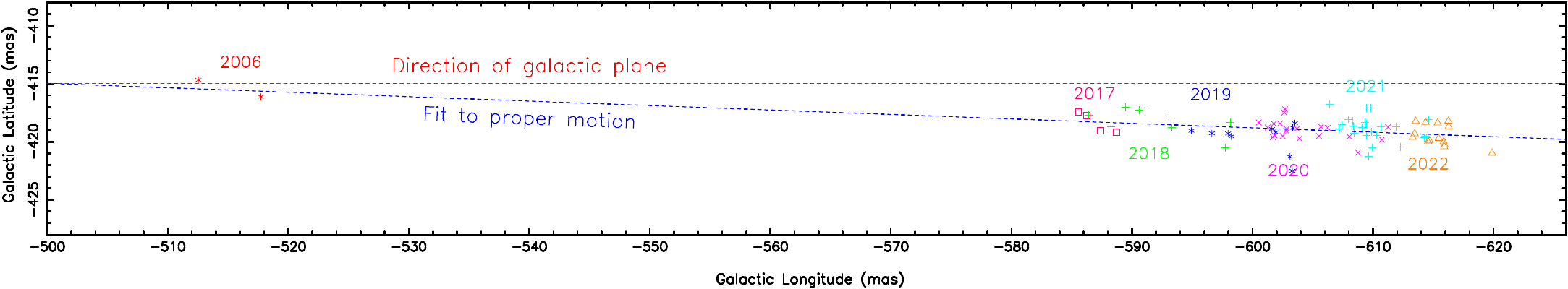}
	\caption {Time series positions of Sgr A* in galactic coordinates,
        offset from -166\arcsec~ galactic latitude, -200\arcsec~ galactic longitude.
       The global solution fit to proper motion is given by the blue 
       dashed line. The direction of the galactic plane is shown by the
       red dashed line.}
       \label{fig:latlon}
\end{figure*}

\begin{figure}[h!]
       \centering
       \includegraphics[width=.45\textwidth]
       {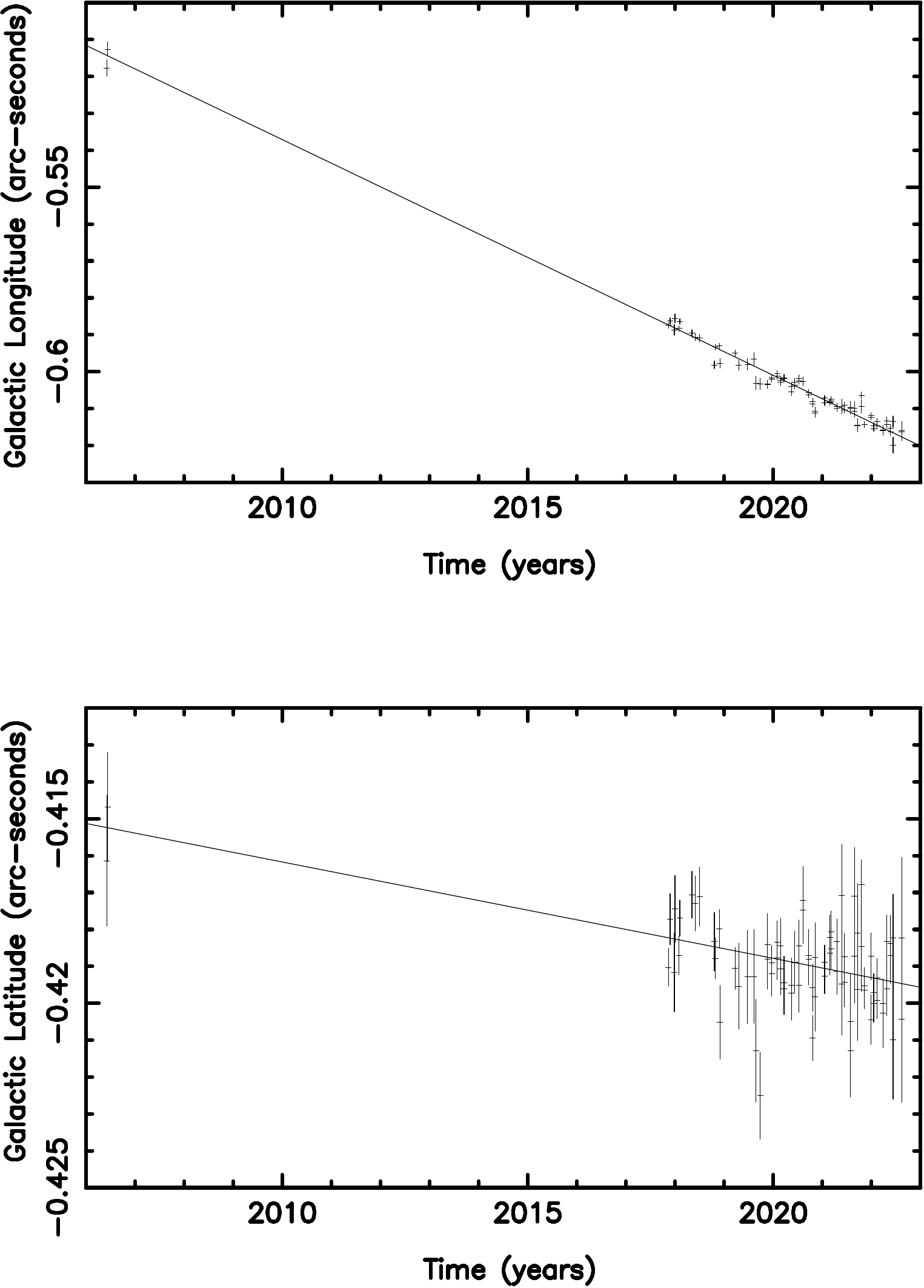}
       \caption {Time series positions of Sgr A* in galactic coordinates from 
	2006-2022, offset from -200\arcsec~ in longitude and -166\arcsec~ in 
	latitude. }
       \label{fig:galslope}
\end{figure}

Transforming the proper motion vectors from the global solution into the 
galactic coordinate system, we get the following proper motion of Sgr A*
in galactic longitude and latitude: \\
\indent $\mu_l$ = -6.396 $\pm$ 0.071 mas ${\rm yr}^{-1}$ \\
\indent $\mu_b$ = -0.239 $\pm$ 0.045 mas ${\rm yr}^{-1}$ \\
To illustrate this graphically, we show in Figure~\ref{fig:latlon} the 
time series positions in galactic coordinates (offset from 
\mbox{-166\arcsec} galactic latitude and \mbox{-200\arcsec} galactic longitude). 
The blue dashed line 
in Figure~\ref{fig:latlon} shows the proper motion direction from the global 
solution, while the red dashed line is parallel to the galactic plane.
And in Figure~\ref{fig:galslope}, we plot the galactic longitudes and 
latitudes (also offset as in Figure~\ref{fig:latlon}) separately versus time.

Thus we see proper motion of -6.396 mas ${\rm yr}^{-1}$ in the galactic plane
and -0.239 mas ${\rm yr}^{-1}$ perpendicular to the galactic plane. If attributed to 
motion of the Sun around the galaxy, we can estimate
the solar velocities in the direction of galaxy rotation and perpendicular
to the plane of the galaxy. 
Using a recent value for the Sun's distance to Sgr A* of 8.178 $\pm$ .022 kpc 
\citep{gravity2019}, we get a tangential solar velocity of 248.0 $\pm$ 2.8 
km ${\rm s}^{-1}$ in 
the galactic plane and 9.3 $\pm$ 1.9 \mbox{km ${\rm s}^{-1}$} out-of-the plane towards
the north galactic pole. This out-of-the plane motion is somewhat larger,
but not significantly different, from a recent value of 7.26 $\pm$ 0.36 
km ${\rm s}^{-1}$ obtained by \citet{ding2019}.

The Sun's tangential motion is a combination of the circular velocity
of the local standard of rest (LSR) and the Sun's peculiar tangential
velocity with respect to the LSR. Several recent studies of the Sun's peculiar 
motion all indicate that the Sun is moving faster around the galaxy than the LSR,
by typically 3 to 12 km ${\rm s}^{-1}$ \citep{dehnen1998, ding2019, golubev2013,
huang2015, schonrich2010}. Of interest for ICRF work is what this could imply
for estimating the galactic aberration constant. Subtracting the various estimated
peculiar velocities, our results would support a galactic aberration constant of
between 4.8 and 5.1 $\mu$as ${\rm yr}^{-1}$. This is somewhat less than the
5.8 $\pm$ 0.3 $\mu$as ${\rm yr}^{-1}$ of \citet{macmillan2019} which was used for
ICRF3 and more in agreement with the 5.05 $\pm$ 0.35 $\mu$as ${\rm yr}^{-1}$
value estimated from Gaia data \citep{gaia2021}.

\section{Conclusions}
We have determined a precise position for Sgr A* as a function of time 
in the frame of the IAU's official celestial reference frame, ICRF3.
Our measurements of its position and proper motion over a 16 year span were
made using absolute VLBI astrometry at 24 GHz on the VLBA and are
independent of previous measurements that used relative 
astrometry. When expressed in galactic coordinates, the proper motion
measured in this study, if attributed to circular motion of the Sun around
the galactic center, indicates solar motion of 248.0 $\pm$ 2.8 km ${\rm s}^{-1}$
in the galactic plane and 9.3 $\pm$ 1.9 km ${\rm s}^{-1}$ towards the north 
galactic pole. 

The VLBA observations used here are part of an ongoing 
collaboration between personnel at the
U.S. Naval Observatory, the South African Radio Astronomy Observatory and
the Jet Propulsion Laboratory. Future updates of this work can be expected to
improve the precision of these measurements.
\\

%%%%%%%%%%%%%%%%%%%%%%%%%%%%%%%%%%%%%%%%%%%%%%%%%%
%\acknowledgments
\section*{Acknowledgements}

The authors gratefully acknowledge use of the Very Long Baseline Array
since 2017 under the US Naval Observatory's time allocation. This work 
supports USNO's ongoing research into the celestial reference frame and 
geodesy. The VLBA is operated  
by the National Radio Astronomy Observatory (NRAO), a  
facility of the National Science Foundation and
operated under cooperative agreement by Associated Universities, Inc.
This work made use of the Swinburne University of Technology software
correlator, DiFX, developed as part of the Australian Major National 
Research Facilities Programme and operated under license. For a 
description of the DiFX correlator, see \citet{deller2011}.
Part of this research was supported by the South African Radio Astronomy 
Observatory (SARAO,) a facility of the National Research Foundation (NRF)
of South Africa.
Part of this research was carried out at the Jet Propulsion Laboratory,
California Institute of Technology, under a contract with the National 
Aeronautics and Space Administration (80NM0018D0004); Copyright 2022,
All Rights Reserved.



\end{document}